\documentclass[3p]{elsarticle}
\usepackage{amsmath}
\usepackage{hyperref}
\usepackage{graphicx}
\usepackage{dcolumn}
\usepackage{multirow}

\begin{document}
\begin{frontmatter}

\title{Universality class of the percolation in two-dimensional lattices with distortion}

\author[mymainaddress]{Hoseung Jang}
\author[mymainaddress]{Unjong Yu\corref{mycorrespondingauthor}}
\cortext[mycorrespondingauthor]{Corresponding author}
\ead{uyu@gist.ac.kr}
\address[mymainaddress]{Department of Physics and Photon Science, Gwangju Institute of Science and Technology, Gwangju 61005, South Korea}

\begin{abstract}

Mitra et al. [Phys. Rev. E 99 (2019) 012117] proposed a new percolation model that includes distortion in the square lattice and concluded that it may belong to the same universality class as the ordinary percolation. But the conclusion is questionable since their results of critical exponents are not consistent. In this paper, we reexamined the new model with high precision in the square, triangular, and honeycomb lattices by using the Newman-Ziff algorithm. Through the finite-size scaling, we obtained the percolation threshold of the infinite-size lattice and critical exponents ($\nu$ and $\beta$). Our results of the critical exponents are the same as those of the classical percolation within error bars, and the percolation in distorted lattices is confirmed to belong to the universality class of the classical percolation in two dimensions.
\end{abstract}


\end{frontmatter}

\section{Introduction}

Percolation \cite{stauffer2014introduction} is one of the most important models in statistical physics since its mathematical model was proposed by Broadbent and Hammersley in 1957 \cite{broadbent1957percolation}. It has been applied to various fields of physics such as conduction \cite{kirkpatrick1973percolation}, magnetic materials \cite{dotsenko1993critical}, metal-insulator transition \cite{ball1994percolative}, and polymer gelation \cite{coniglio1979site}. It has been also studied outside of physics in relation to forest fires \cite{henley1993statics}, epidemics \cite{cardy1985epidemic,moore00epidemic,davis08epidemic}, and network robustness \cite{cohen2000resilience, moreira2009make}, etc.

One of the simplest models of the percolation is site percolation. In this model, every site of the regular lattice is occupied randomly and independently with an occupation probability $p$. If two nearest-neighboring sites are occupied, they form a cluster. Another simple percolation model is bond percolation, in which empty bonds between neighboring sites on the lattice are occupied randomly with probability $p$. The two models can be integrated resulting in site-bond percolation model, where the sites are occupied with probability $p_s$ and the bonds between two occupied nearest-neighboring sites are occupied with probability $p_b$. Although the three models show differences in details, they show similar behaviors. The size of the largest cluster shows a very steep but continuous increase at a certain value of $p = p_c$, which is called a percolation threshold. The values of the percolation threshold for various lattices and dimensions have been obtained numerically and analytically \cite{suding1999site,wang2013bond,jacobsen2014high,ziff1997determination,van1997percolation,sykes1964exact,sykes1964critical,malarz2015simple}. Near the point of the transition, some quantities show universal behaviors that can be described by the critical exponents. A collection of models that have the same critical exponents is called a universality class. The three models are known to belong to the same universality class \cite{stauffer2014introduction, PhysRevB.45.1023}.

A few interesting variation models of percolation were proposed over the years. Kundu and Manna introduced a percolation model with an additional source of disorder \cite{kundu2016percolation}. The sites are occupied with a disc with random radii and the occupation of the bond between two nearest neighbors is determined by the sum or product of radii of the two discs in the sites. A site with a large disc tends to have occupied bonds with its neighbors, and so the occupations of the bonds that are connected to the same site are positively correlated. Mitra et al. introduced another percolation with a disorder \cite{mitra2019percolation}. In their model, the position of a site on a lattice is randomly shifted on the certain domain that is controlled by an adjustable parameter, and two sites are connected when their Euclidean distance is smaller than a certain distance; thus, the occupations of the bonds that are connected to the same site are negatively correlated. Both models were concluded to belong to the same universality class as the ordinary percolation model based on the values of their critical exponents. However, the conclusion of Mitra et al. is questionable because all the values of critical exponent ($\beta/\nu = 0.1143(20)$, $0.1158(26)$, and $0.1158(30)$ for three kinds of distortion) show systematic deviation from $\beta/\nu = 5/48 \approx 0.1042$ \cite{stauffer2014introduction} of the ordinary two-dimensional percolation by four or five times of the error bars. Therefore, the conclusion should have been that the model with distortion belongs to a different kind of the universality class from their results. 

In this work, we reexamined the model of Mitra et al. in the square lattice and extended this model to the triangular and honeycomb lattices. We calculated the percolation threshold of the infinite-size lattices and critical exponents ($\nu$ and $\beta$) for the three lattices to check whether this model belongs to the same universality class as the ordinary percolation model in two dimensions.

\begin{figure*}
\includegraphics[width=1\columnwidth]{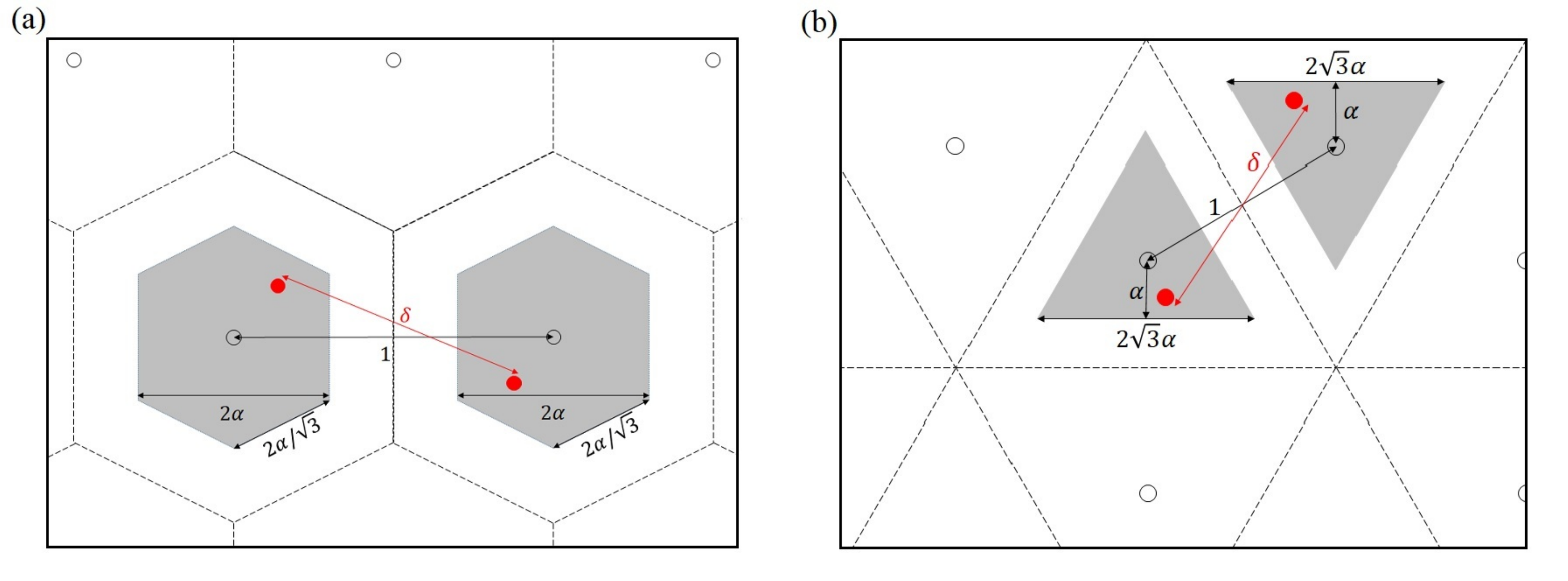}
\caption{\label{fig:fig1} A domain (in grey) of the random uniform sampling for the distortion (a) in the triangular lattice and (b) in the honeycomb lattice with regular lattice points (empty circles). A possible configuration of two neighboring sites with distortion through uniform random sampling is also shown with red circles. The dashed lines represent the Wigner-Seitz cell of the lattice, which is the same as the random uniform sampling domain with $\alpha = 0.5$.}
\end{figure*}

\section{Methods}

A distorted square lattice was generated by the method of Ref.~\cite{mitra2019percolation}: each site in the square lattice is shifted by $r_x$ and $r_y$ in the horizontal and vertical directions, respectively. The two values of $r_x$ and $r_y$ are chosen from $[-\alpha, \alpha]$ uniformly at random, where the distortion parameter $\alpha$ is not larger than 0.5 when the bond length of the undistorted lattice is 1. Therefore, the site is placed uniformly at random within a square of length $2\alpha$ centered at the regular lattice point. For the extreme case of $\alpha = 0.5$, the square becomes the Wigner-Seitz cell and the squares fill entire space. When we generate distorted triangular and honeycomb lattices, we need to modify the method used in the square lattice in order to avoid over-distortion.
Just like in the square lattice, we set up the sampling domain to be the Wigner-Seitz cell for $\alpha = 0.5$ in the triangular and honeycomb lattices. For $\alpha < 0.5$, the sampling domain is a smaller isomorphic domain, which is a regular hexagon of side length $2\alpha/\sqrt{3}$ and a regular triangle of side length $2\sqrt{3}\alpha$ for the triangular and honeycomb lattices, respectively. (See Fig.~\ref{fig:fig1}.) A new site of the distorted lattice is chosen uniformly at random within the sampling domain. Two occupied nearest neighbors are set to be connected if and only if their Euclidean distance is smaller than the connection threshold $d$.

In this work, the Newman-Ziff algorithm, which is much more efficient than brute-force methods when the percolation threshold is not known in advance \cite{newman2000efficient, newman2001fast, choi2019newman}, is used. In this algorithm, a set of average values $\langle Q(n)\rangle$ is obtained first, where $Q(n)$ is any physical quantity such as the size of the largest cluster for a fixed number of occupied sites $n$. A value of $\langle Q(p)\rangle$ as a function of occupation probability $p$ can be calculated by the binomial transformation:
\begin{equation}
\langle Q(p) \rangle = \sum^{N}_{n=0}\frac{N!}{n!(N-n)!}p^{n}(1-p)^{N-n}\langle Q(n) \rangle,
\end{equation}
where $N$ is the total number of sites in the lattice. Therefore, we can get $\langle Q(p)\rangle$ for an arbitrary value of $p$ by getting $\langle Q(n)\rangle$ once. Other algorithms than the Newman-Ziff algorithm shows quantities only for one specified value of $p$ at one time. They have a limitation that they can give quantities only on a finite number of $p$ values. In contrast, the Newman-Ziff algorithm can give a continuous set of quantities on the entire domain $0 \le p \le 1$  by the binomial transformation. Another advantage of the Newman-Ziff algorithm is that we can get derivatives of $\langle Q(p)\rangle$ without numerical differentiation by differentiating the above equation analytically \cite{newman2001fast, choi2019newman}:
\begin{equation}
\frac{d\langle Q(p) \rangle}{dp} = \sum^{N}_{n=0}\frac{N!}{n!(N-n)!}p^{n-1}(1-p)^{N-n-1}(n-Np)\langle Q(n) \rangle.
\end{equation}
It prevents inevitably large error from the numerical differentiation. Thus, the Newman-Ziff algorithm gives us more precise results than other algorithms in general.
\begin{figure*}
\includegraphics[width=1\columnwidth]{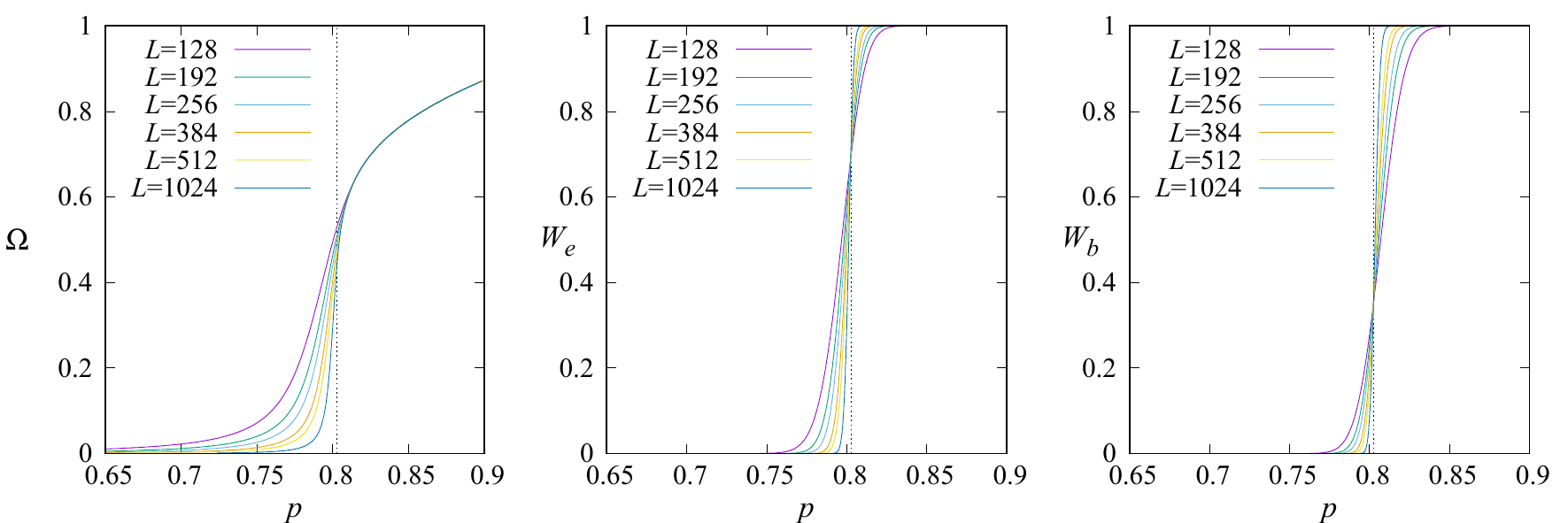}
\caption{\label{fig:fig2}Plots of the order parameter $\Omega(p,L)$,  wrapping probability in either direction $W_e (p,L)$, and that in both directions $W_b (p,L)$ in the square lattice with $\alpha = 0.2$ and $d = 1.1$ for various lattice sizes $L=$ 128, 192, 256, 384, 512, and 1024. The vertical dotted lines represent our estimation of the percolation threshold of the infinite-size lattice.}
\end{figure*}

\section{Results and discussion}

The order parameter $\Omega(p,L)$ is defined as 
\begin{equation}
\Omega(p,L)=\langle S_{\mathrm{max}} \rangle/N
\end{equation}
where $S_{\mathrm{max}}$ is the size of the largest cluster and $N$ is the total number of sites in the lattice, which is $N = L^2$ for the square and triangular lattices and $N = 2L^2$ for the honeycomb lattice.  Figure~\ref{fig:fig2} shows the order parameter $\Omega(p,L)$, wrapping probability in either direction $W_e(p,L)$, and that in both directions $W_b(p,L)$ in the distorted square lattice for various lattice sizes $L=$ 128, 192, 256, 384, 512, and 1024. They exhibit a continuous percolation transition just like the ordinary percolation model. These behaviors are also observed in the distorted triangular and honeycomb lattices.

\begin{figure*}
\includegraphics[width=1\columnwidth]{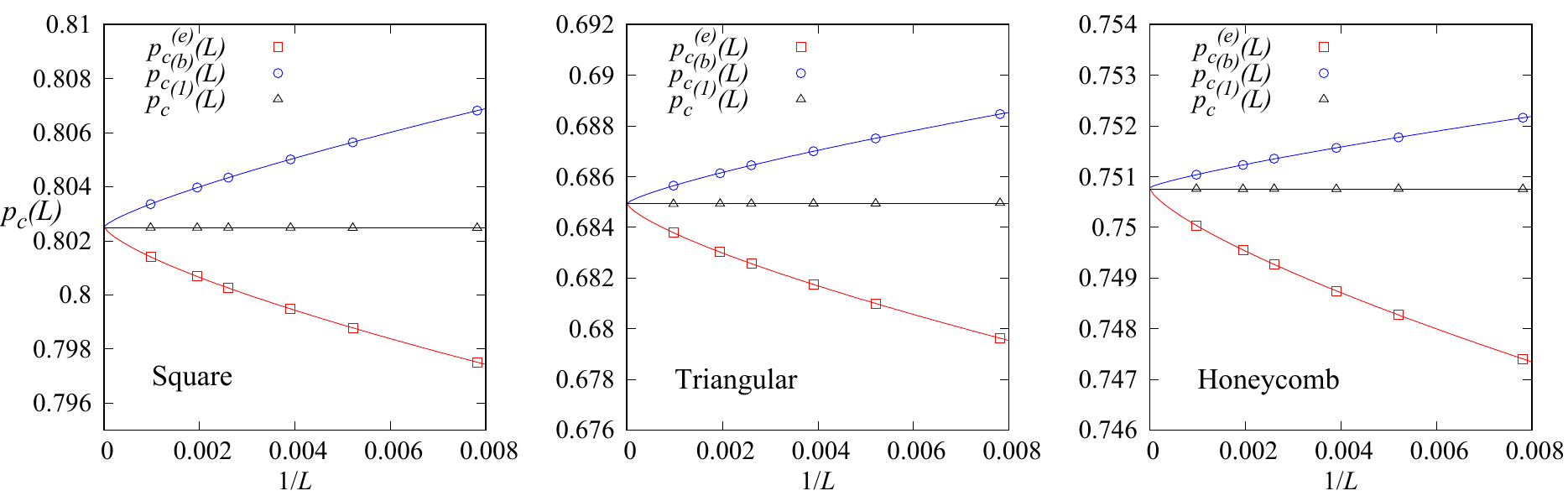}
\caption{\label{fig:fig3}The three values of the percolation threshold of the finite-size lattice $p_{c}^{(e)}(L)$ (square), $p_{c}^{(b)}(L)$ (circle), and $p_{c}^{(1)}(L)$ (triangle) obtained from the maximum of $dW_e(p,L)/dp$, $dW_b(p,L)/dp$, and $W_1(p,L)=[W_e (p,L)-W_b (p,L)]$ (a) in the distorted square lattice, (b) in the distorted triangular lattice, and (c) in the distorted honeycomb lattice. Distortion parameter and connection threshold are $\alpha=0.2$ and $d=1.1$ for the square and triangular lattices, and $\alpha=0.1$ and $d=1.15$ for the honeycomb lattice. The error bar is smaller than the symbol size. The solid lines are fitting curves according to the finite-size scaling relations.}
\end{figure*}

The percolation threshold of the infinite-size lattice $p_{c}^{\infty}$ is obtained by the finite-size scaling $[p_c (L)-p_{c}^{\infty} ] \sim L^{-a}$, where $p_c(L)$ is a percolation threshold of the finite-size lattice and $a$ is a fitting parameter. The value of the fitting parameter $a$ depends on the type of the quantity, the percolation mechanism, and the lattice \cite{ziff2002convergence}. The three values of the percolation threshold of the finite-size lattice $p_c(L)$ are obtained from the maximum point of $dW_e (p,L)/dp$, $dW_b (p,L)/dp$, and $W_1(p,L)=[W_e (p,L)-W_b (p,L)]$ for various lattice sizes in each lattice.
$W_1(p,L)$ is the wrapping probability in any one direction but not in the other direction.
Figure~\ref{fig:fig3} shows the finite-size scaling behavior of the percolation threshold of the finite-size lattice $p_c(L)$ and estimation of the percolation threshold of the infinite-size lattice $p_c^{\infty}$ by fitting curves according to the scaling relations in the square, triangular, and honeycomb lattices. From the fitting, $p_c^{\infty}=0.80251(6)$ for the square lattice with $\alpha = 0.2$ and $d = 1.1$, $p_c^{\infty}=0.68494(4)$ for the triangular lattice with $\alpha = 0.2$ and $d = 1.1$, and $p_c^{\infty}=0.75077(3)$ for honeycomb lattice with $\alpha = 0.1$ and $d = 1.15$. (There is no percolation in the honeycomb lattice with $\alpha = 0.2$ and $d = 1.1$.) These threshold values are used in the calculation of the critical exponents.

\begin{figure*}
\includegraphics[width=1\columnwidth]{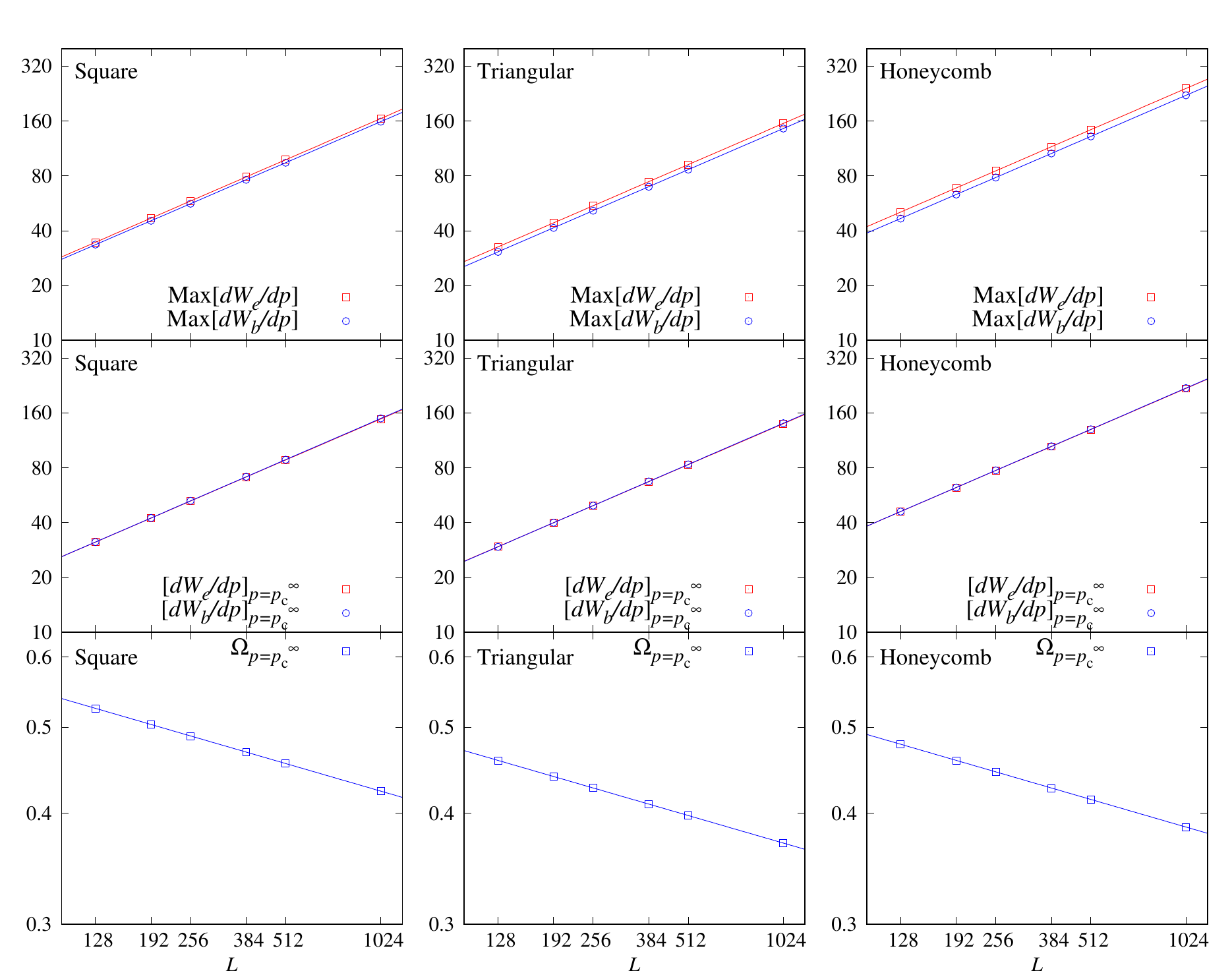}
\caption{\label{fig:fig4}Values of $dW_e(p,L)/dp$ and $dW_b(p,L)/dp$ at the maximum point (upper), those of at the percolation threshold of the infinite-size lattice (middle), and order parameter $\Omega(p,L)$ at the percolation threshold of the infinite-size lattice (lower) as a function of the system size $L$ for the distorted square, triangular, and honeycomb lattices. Parameters $\alpha$ and $d$ are the same as Fig.~\protect\ref{fig:fig3}. The error bar is smaller than the symbol size. All plots are in log-log scale. The solid lines are fitting curves according to the scaling relations.}
\end{figure*}

Critical exponents $\nu$ and $\beta$ can be estimated from the scaling relations \cite{lobb1980monte,martins2003percolation, choi2019newman}
\begin{equation}
\text{Max}\left[\frac{dW(p,L)}{dp}\right]  \sim L^{1/\nu}, \quad \left[\frac{dW(p,L)}{dp}\right]_{p=p_{c}^{\infty} } \sim L^{1/\nu},\quad \text{and}\quad\Omega(p=p_c^{\infty},L) \sim L^{\beta/\nu},
\end{equation}
where $W(p,L)$ stands for both $W_e(p,L)$ and $W_b(p,L)$. We calculated values of $dW_e(p,L)/dp$ and $dW_b(p,L)/dp$ at the maximum point, those at the percolation threshold of the infinite-size lattice, and $\Omega(p,L)$ at the percolation threshold of the infinite-size lattice in the square, triangular, and honeycomb lattices for various lattice sizes.
Figure~\ref{fig:fig4} shows the scaling behaviors of these values. Since the plots are in log-log scale, the slope of the straight lines in the upper and middle plots is $1/\nu$ and the slope of the straight lines in the lower plots is $\beta/\nu$. In the estimation of the error of critical exponents obtained from the quantities at the threshold point, however, we should consider another source of the error from an inaccurate threshold point \cite{yu2017phase}. We observed that the value of $\beta/\nu$ changes by about 1.83\% when $p$ changes by 0.01\% from $p_c^{\infty}$ in the triangular lattice. Since the uncertainty of $p_c^{\infty}$ in the triangular lattice is 0.00004 (about 0.0058\% of $p_c^{\infty}$), the maximum error of $\beta/\nu$ due to the uncertainty is about 1.06\% of its value. By the same way, we estimated the error of $1/\nu$ obtained from $\left[dW(p,L)/dp\right]_{p=p_{c}^{\infty} }$. All these considerations were also made on the other two lattices. The estimated values of $1/\nu$ and $\beta/\nu$ are summarized in Table~\ref{tab:table1}. The comparison with the critical exponents of the ordinary percolation model ($1/\nu = 0.75$ and $\beta/\nu = 5/48 \approx 0.1042$ in two dimensions \cite{stauffer2014introduction}) shows that the percolation model in distorted lattices belongs to the universality class of the ordinary percolation model in two dimensions.

\begin{table}
\caption{\label{tab:table1}
The estimated values of the critical exponents ($\nu$ and $\beta$) of the percolation in the square lattice with $\alpha = 0.2$ and $d = 1.1$, triangular lattice with $\alpha = 0.2$ and $d = 1.1$, and honeycomb lattice with $\alpha = 0.1$ and $d = 1.15$.
}
\resizebox{\linewidth}{!}{
\begin{tabular}[]{|c|c|c|c|l|c|c|}
\hline
\multirow{2}{*}{Lattice} & \multicolumn{5}{c|}{$1/\nu$ }                                                                                                                                                                              & $\beta/\nu$                \\ \cline{2-7} 
                         & $\text{Max}\left[dW_{e}(p,L)/dp\right]$ & $\text{Max}\left[dW_{b}(p,L)/dp\right]$ & \multicolumn{2}{c|}{$\left[dW_{e}(p,L)/dp\right]_{p=p_{c}^{\infty} }$} & $\left[dW_{b}(p,L)/dp\right]_{p=p_{c}^{\infty} }$ & $\Omega(p=p_c^{\infty},L)$ \\ \hline
Square                   & 0.752(2)                               & 0.748(2)                               & \multicolumn{2}{c|}{0.747(7)}                                         & 0.752(5)                                         & 0.103(2)                      \\
Triangular               & 0.751(2)                               & 0.750(1)                               & \multicolumn{2}{c|}{0.747(6)}                                         & 0.751(3)                                         & 0.103(1)                      \\
Honeycomb                & 0.751(1)                               & 0.749(1)                               & \multicolumn{2}{c|}{0.748(5)}                                         & 0.750(2)                                         & 0.103(2)                      \\ \hline
\end{tabular}
}
\end{table}

\section{Conclusions}

We reexamined percolation on the square lattice with distortion proposed by Mitra et al. with high precision using the Newman-Ziff algorithm. In addition, we extended this distortion model to the triangular and honeycomb lattices. A distortion is given to the lattice through random shifts of a position of each regular lattice site within an isomorphic domain of the Wigner-Seitz cell whose size is controlled by the parameter $\alpha$. Through the finite-size scaling, we calculated the percolation threshold $p_c^{\infty}$ and critical exponents $\beta$ and $\nu$ in the square, triangular, and honeycomb lattices. Contrary to the previous results of Mitra et al., our results of critical exponents are the same as those of the ordinary percolation within error bars, implying that this model almost certainly belongs to the same universality class as the ordinary percolation in two dimensions.

\section*{Acknowledgments}
This work was supported by GIST Research Institute (GRI) grant funded by the GIST in 2019.


\bibliography{Ref}

\end{document}